\documentclass[conference]{IEEEtran}
\IEEEoverridecommandlockouts
\usepackage[dvipsnames]{xcolor}
\usepackage{cite}
\usepackage{amsmath,amssymb,amsfonts}
\usepackage{graphicx}
\usepackage{textcomp}
\usepackage{xcolor}
\usepackage{algorithm} 
\usepackage{algpseudocode}
\usepackage{hyperref}
\usepackage{url}
\def\BibTeX{{\rm B\kern-.05em{\sc i\kern-.025em b}\kern-.08em
    T\kern-.1667em\lower.7ex\hbox{E}\kern-.125emX}}
\begin{document}

\title{Hybrid Variational Quantum-Classical Framework with Adaptive Weighting and Efficiency Assessment 
}

\author{\IEEEauthorblockN{Dilli Hang Rai}
\IEEEauthorblockA{{Institute of Science and Technology} \\
{Tribhuvan University, Kritipur, Nepal}\\
{dillihangrae@gmail.com}
}}

\maketitle

\begin{abstract}
Hybrid quantum-classical neural networks have emerged as a promising approach
for leveraging quantum computing in machine learning
while mitigating current hardware limitations.
This paper presents Sim-HVQC,
a hybrid Deep Quantum Neural Network that couples an adaptive,
parameter-free SimAM weighting module with classical feature extraction
to preserve class-discriminative information prior to encoding into a
Variational Quantum Circuit (VQC).
Previous studies are restricted to binary classification 
\cite{chen2020hybridquantumclassicalclassifierbased}\cite{2021_chen_anendtoend}\cite{2024_hybrid_quantum_classical}\cite{Long2025Hybrid}\cite{2023_dasvariationalquantumneuralnetworks}.
In contrast, the proposed framework is trained and evaluated on various
multi-class datasets(MNIST, KMNIST, Fashion-MNIST, and EMNIST).
The framework further demonstrates reproducibility, parameter efficiency, 
and interpretability through multi-seed evaluation, parameter analysis, and latent/quantum feature inspection. 
The source code is publicly available at {\normalfont \url{https://github.com/Dilli822/SimAM-HVQC}.}
\end{abstract}


\section{Introduction}
\label{sec:introduction}

Quantum machine learning offers one such direction and hybrid quantum neuralnetworks, 
which combine classical neural-network components with quantum information processing units, 
have emerged as a practical framework for near-term quantum technologies \cite{2026_monbroussouhybridquantumneuralnetworks}.
Existing HQNN studies are commonly evaluated on binary-classification settings or a single benchmark dataset, while reproducibility analysis, efficiency assessment, and model interpretability remain comparatively underexplored in hybrid quantum-classical image-classification frameworks.
In particular, \cite{2023_dasvariationalquantumneuralnetworks} focused on binary classification, 
while \cite{2022_choe_continuousvariablequantummnist} performed multi-class classification using a single dataset. 
Both studies reported only training accuracy, without validation or test results. 
Although \cite{Senokosov_2024} achieved impressive accuracy, its evaluation was limited to binary classification. 
Similarly, \cite{2024_HQNN_dataset_experiment} conducted experiments across datasets but remained restricted 
to binary classification. 
These previous studies highlight the need for dataset-level experiments involving a larger number of multi-class, 
multi-dataset evaluations with empirical validation and test performance. 

\begin{figure}[htbp]
    \centering
    \includegraphics[width=\columnwidth]{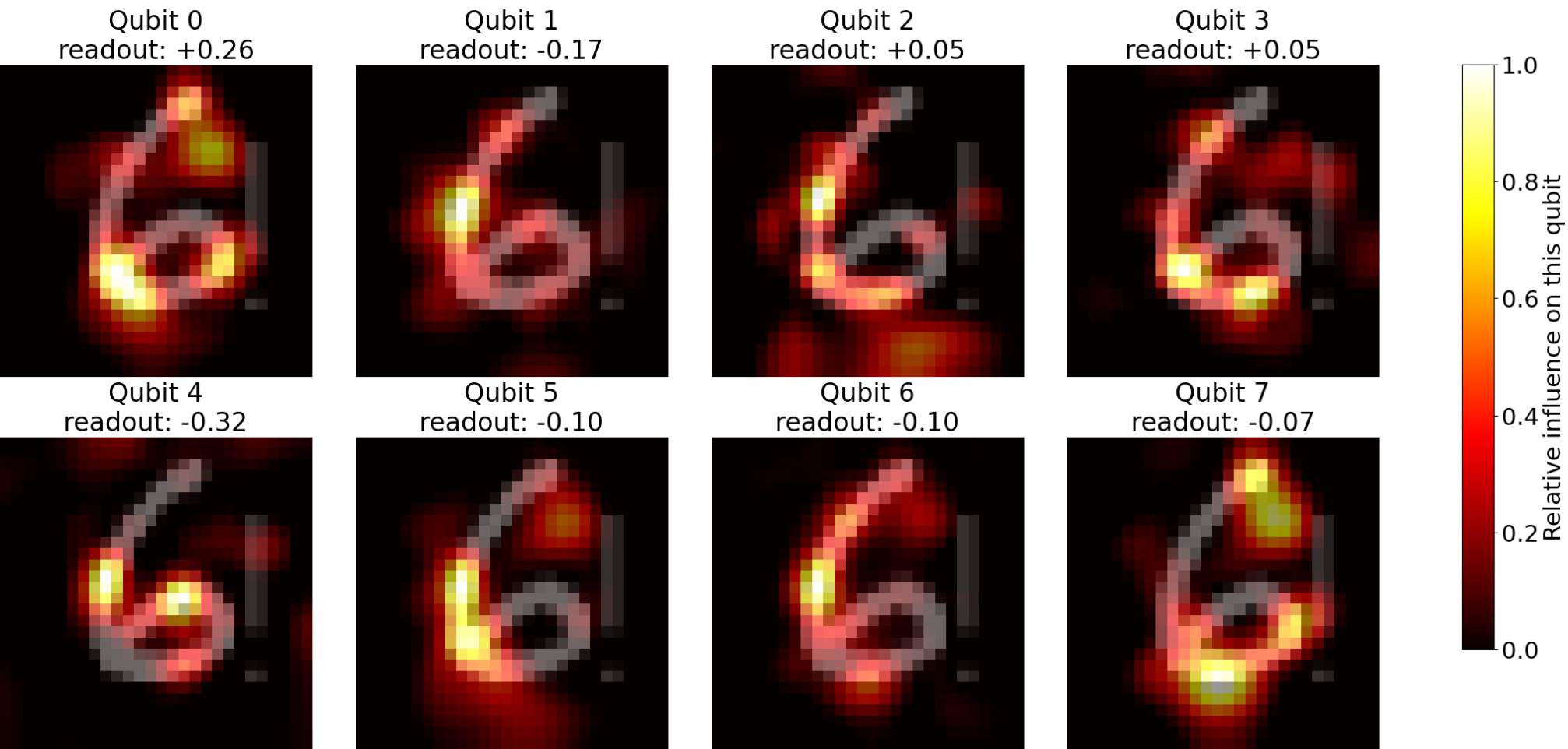}
    \caption{Occlusion-based qubit sensitivity maps for a sample input. The varying sensitivity patterns across qubits indicate that different quantum measurements capture complementary information from the compressed latent representation}
    \label{fig:occulsion_6}
\end{figure}

The study \cite{2024_zaman_comparativeanalysishybridquantumclassical} demonstrated that different hybrid quantum-classical 
architectures exhibit varying performance characteristics under different circuit configurations.
Therefore, continued exploration of improved hybrid quantum-classical architectures remains necessary. Motivated by the growing interest in practical hybrid quantum 
neural networks and the need to explore improved hybrid architectures \cite{2026_monbroussouhybridquantumneuralnetworks}, 
This paper proposes SimAM-HVQC, a parameter-efficient hybrid quantum-classical model for multi-class image classification that adopts an alternative approach by incorporating a parameter-free attention mechanism prior to quantum encoding.

The contributions of this paper are listed below:
\begin{itemize}
    \item Proposed a hybrid Deep Quantum Neural Network (DQNN) that integrates a SimAM attention-enhanced classical feature extractor with a Variational Quantum Circuit (VQC).


    \item Evaluated the framework on multiple standard benchmarks with varying numbers of classes:MNIST,Fashion-MNIST(F-MNIST), 
          KMNIST and EMNIST.
          
    \item Assessed model reproducibility through multi-seed experiments, analyzed performance under varying training-data sizes, and provided a lightweight interpretability analysis of latent feature representations and quantum measurement outputs.
\end{itemize}

\section{Methodology}
\label{sec:methodology}
This section describes the dataset preparation, classical
feature extraction, and the proposed hybrid quantum-classical
architecture, along with the corresponding algorithm.
\subsection{Dataset}
\label{subsec:dataset}
The proposed framework was evaluated on MNIST, Fashion-MNIST, KMNIST and EMNIST  using a standard 70:15:15 train-validation-test split.
MNIST, Fashion-MNIST, and KMNIST are 10-class, 28×28 grayscale datasets; EMNIST Balanced has 47 classes of 28×28 grayscale character images.

\subsection{SimAM Attention} 
\label{subsec:simam_attention} 

Before dimensionality reduction, the input representation is refined using the parameter-free Simple Attention Module (SimAM). 
Given an input feature map \begin{equation} X \in \mathbb{R}^{C\times H\times W}, \end{equation} 
SimAM computes neuron-wise attention coefficients using an energy-based formulation \cite{yang2021simam}. 
The channel-wise mean and variance are 
\begin{equation} \mu=\frac{1}{M}\sum_{i=1}^{M}x_i, \qquad \sigma^2=\frac{1}{M}\sum_{i=1}^{M}(x_i-\mu)^2, \end{equation} 
where \(M=H\times W\). For neuron \(x_i\), 
the inverse energy is 
\begin{equation} E_i^{-1} = \frac{(x_i-\mu)^2} {4(\sigma^2+\lambda)} +\frac{1}{2}, \end{equation} where \(\lambda\) is a regularization parameter. The attention coefficient is obtained as \begin{equation} A_i=\sigma(E_i^{-1}), \end{equation} where \(\sigma(\cdot)\) denotes the sigmoid function. The attention-refined representation is then computed by \begin{equation} X_{\mathrm{SimAM}} = X \odot A, \end{equation} where \(\odot\) denotes element-wise multiplication. Consequently, activations exhibiting greater distinguishability from the underlying channel distribution receive larger attention coefficients. The resulting representation \(X_{\mathrm{SimAM}}\) is forwarded to the classical feature extractor for subsequent compression and quantum encoding.

\subsection{Classical Neural Network} 
\label{subsec:classical_neural_network} 
The classical neural network $f_{\mathrm{feat}}(\cdot)$ functions as 
a learnable feature extractor and representation learner, 
transforming the SimAM-enhanced input $\mathbf{X}_a$ into a discriminative 
$Q$-dimensional latent embedding $\mathbf{F} = f_{\mathrm{feat}}(\mathbf{X}_a) \in \mathbb{R}^{Q}$ 
that supports downstream quantum processing and classification.

\subsection{Quantum Processing}
\label{subsec:quantum_processing}

The compressed latent representation $F \in \mathbb{R}^{Q}$ is encoded into a variational quantum circuit (VQC), enabling feature transformation in a high-dimensional Hilbert space. The encoded quantum state evolves as \begin{equation} |\psi\rangle = U(\omega,\theta)|0\rangle^{\otimes Q}, \end{equation} where $\theta=\pi\tanh(F)$ and $U(\cdot)$ denotes the trainable quantum circuit. The VQC exploits superposition and entanglement to capture complex feature interactions, producing quantum-enhanced representations that are difficult to express through the preceding low-dimensional classical mapping alone. Quantum features are extracted through Pauli-$Z$ expectation measurements, \begin{equation} z_i=\langle\psi|Z_i|\psi\rangle, \qquad i=1,\ldots,Q, \end{equation} and forwarded to the final classifier. In the proposed framework, the quantum module employs $8$ qubits and $6$ strongly entangling layers, resulting in $144$ trainable quantum parameters.

\begin{figure*}[t]
    \centering
    \includegraphics[width=\textwidth,height=\textheight,keepaspectratio]{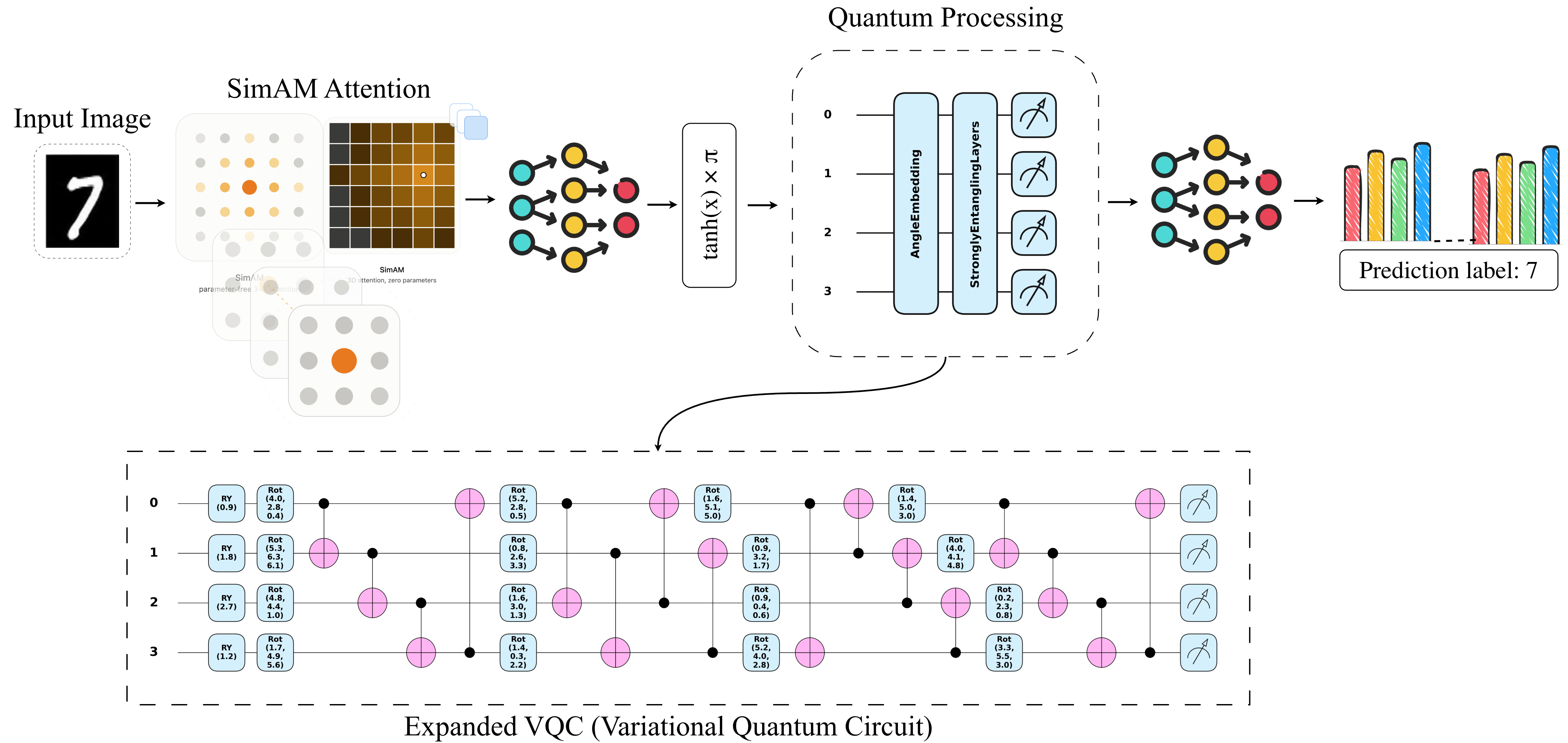}
    \label{fig:vqc}
    \caption{Architecture of the proposed SimAM-enhanced hybrid variational quantum-classical neural network 
    (SIMAM-HVQC). The input image is first processed by the parameter-free SIMAM attention module to enhance 
    informative feature responses. The refined feature map is flattened and passed through a classical 
    fully connected feature extractor, reducing the original $784$-dimensional representation to $256$, $128$, 
    and finally $Q$ quantum-compatible features. The resulting features are scaled 
    using $\boldsymbol{\theta}=\pi\tanh(\mathbf{F})$ and encoded into the quantum register 
    through $R_Y$-based AngleEmbedding. The variational quantum component consists of $Q$ qubits 
    and $L$ trainable Strongly Entangling Layers, followed by Pauli-$Z$ expectation-value 
    measurements to produce $Q$ quantum features. These measurements are subsequently processed 
    by a classical fully connected classifier $Q\rightarrow H\rightarrow C$, where $H$ denotes the 
    hidden-layer dimension and $C$ the number of target classes, producing the final 
    class prediction $\hat{Y}$. The lower panel provides the corresponding gate-level 
    representation of the variational quantum circuit, illustrating the trainable 
    single-qubit rotations and entangling operations.}
\end{figure*}

\begin{algorithm}[t]
\caption{Proposed SIMAM-HQNN with a VQC }
\label{alg:simam_hqnn}
\begin{algorithmic}[1]
\Require Input image $X$; number of qubits $Q$; number of quantum layers $L$;
number of classes $C$
\Ensure Predicted class scores $\hat{Y}$
\State $X_{a} \gets \operatorname{SIMAM}(X)$
\State $\mathbf{x} \gets \operatorname{Flatten}(X_{a})$
\State $\mathbf{F} \gets
\operatorname{FC}_{784\rightarrow256\rightarrow128\rightarrow Q}(\mathbf{x})$
\State $\boldsymbol{\theta} \gets \pi\tanh(\mathbf{F})$
\State Encode $\boldsymbol{\theta}$ on $Q$ qubits using angle embedding
\State Apply $L$ strongly entangling layers
\State $\mathbf{Z} \gets
\{\langle Z_{1}\rangle,\ldots,\langle Z_{Q}\rangle\}$
\State $\hat{Y} \gets
\operatorname{FC}_{Q\rightarrow64\rightarrow C}(\mathbf{Z})$
\State \Return $\hat{Y}$
\end{algorithmic}
\end{algorithm}

\section{Experimental Setup}
\label{sec:experimental_setup}
This section presents the experimental configuration adopted in the paper. 
Table~\ref{tab:full-config} summarizes the datasets, model architecture, quantum-circuit configuration, and training hyperparameters.

\begin{table}[!t]
\renewcommand{\arraystretch}{1.3}
\caption{Model Configuration, Dataset, and Hyperparameter Summary}
\label{tab:full-config}
\centering
\begin{tabular}{|l|l|}
\hline
\textbf{Parameter} & \textbf{Value} \\
\hline
\multicolumn{2}{|c|}{\textit{Dataset}} \\
\hline
Dataset used             & MNIST, KMNIST, EMNIST, FashionMNIST \\
\hline
Input image dimension    & $28 \times 28$ (grayscale) \\
\hline
Flattened input size     & 784 \\
\hline
Number of classes        & 10, 47 \\
\hline
\multicolumn{2}{|c|}{\textit{Layer Dimensions (In $\rightarrow$ Out)}} \\
\hline
Classical encoder L0     & 784 $\rightarrow$ 256 \\
\hline
Classical encoder L1     & 256 $\rightarrow$ 128 \\
\hline
Classical encoder L2     & 128 $\rightarrow$ 8 \\
\hline
Quantum layer            & 8 $\rightarrow$ 8 (qubits) \\
\hline
Classifier L0            & 8 $\rightarrow$ 64 \\
\hline
Classifier L1            & 64 $\rightarrow$ number of dataset classes \\
\hline
\multicolumn{2}{|c|}{\textit{Training Hyperparameters}} \\ 
\hline 
Random seed & 42, 32, 102, 72, 2 \\ \hline 
Batch size & 64 \\ \hline 
Epochs & 25 \\ \hline 
Learning rate & 0.001 \\ \hline 
Optimizer & Adam \\ \hline 
Loss Function & Weighted Cross-Entropy Loss \\ \hline 
Device & CPU \\ \hline 
Train/Validation/Test Split & 70:15:15 \\ \hline 
Gradient Clipping & 1.0 \\ \hline 
Activation Function & ReLU \\ \hline 
SIMAM Parameter & $\lambda = 10^{-4}$ \\ \hline 
\multicolumn{2}{|c|}{\textit{Quantum Configuration}} \\ \hline 
Quantum Encoding & AngleEmbedding \\ \hline 
Variational Ansatz & Strongly Entangling Layers \\ \hline 
Measurement Observable & Pauli-Z Expectation Values \\ \hline 
Differentiation Method & Backpropagation \\ \hline 
Quantum Simulator & PennyLane 
\texttt{default.qubit} \\ \hline
\end{tabular}
\end{table}

\section{Results}
\label{sec:results-analysis}
This section presents the results obtained from the proposed method.

\begin{table}[htbp]
\caption{Comparison of Proposed Method Performance with the Existing Methods Performance}
\label{tab:sota}
\centering
\tiny
\begin{tabular}{|l|c|c|c|c|c|c|}
\hline
\textbf{Study} &
\textbf{Method} &
\textbf{Qubits} &
\textbf{Dataset} &
\textbf{Classes} &
\textbf{Loss} &
\textbf{Acc. (\%)} \\
\hline

\cite{chen2020hybridquantumclassicalclassifierbased}
& MPS-VQC
& 4
& MNIST
& 2
& 0.3183
& 99.44 \\
\hline

&  
&  
& MNIST
& 4
& -
& 85.14   \\

\cite{2022Multiclassqcnn}
& QCNN
& 12 qubits
& MNIST
& 4
& -
& 90.03 \\

& (Hybrid VQC)
& 
& F-MNIST
& 4
& -
& 85.93 \\
\hline

\cite{zeng2022multi}
& HQNN
& 16
& MNIST
& 10
&  -  
&  89.06 \\
\hline

\cite{Senokosov_2024}
& HQNN
& 5
& MNIST
& 2
&  0.0274  
&  99.21 \\
\hline

\cite{2024_superposition}
& SEQNN
& -
& MNIST
& 10
& -
& 87.56
\\
\hline

\cite{anwar2025}
& Hybrid QCNN
& 8
& MNIST 
& 4
& -
& 88.52
\\

&
& 8
& F-MNIST 
& 4
& -
& 86.55
\\
&
& 8
& MNIST 
& 4
& -
& 93.55
\\
\hline

\textbf{}
& \textbf{}
& \textbf{8}
& \textbf{MNIST}
& \textbf{10}
& \textbf{0.1437 ±  0.013}
& \textbf{97.57 ± 0.05} \\

\textbf{}
& \textbf{}
& \textbf{8}
& \textbf{F-MNIST}
& \textbf{10}
& \textbf{0.3349± 0.0131}
& \textbf{87.98 ± 0.08}  \\

\textbf{Proposed}
& \textbf{SimAM-HVQC}
& \textbf{8}
& \textbf{EMNIST}
& \textbf{47}
& \textbf{0.7212 ± 0.0236}
& \textbf{80.16 ± 0.17}  \\

\textbf{}
& \textbf{}
& \textbf{8}
& \textbf{KMNIST}
& \textbf{10}
& \textbf{0.3545 ± 0.0273}
& \textbf{88.20 ± 0.61}  \\
\hline

\end{tabular}
\end{table}

\begin{table}[htbp]
\caption{Performance at Different Dataset Utilization Levels}
\label{tab:dataset_utilization}
\centering
\tiny
\begin{tabular}{|l|c|c|c|c|c|c|c|c|}
\hline

\textbf{Dataset} &
\multicolumn{2}{c|}{\textbf{KMNIST}} &
\multicolumn{2}{c|}{\textbf{F-MNIST}} &
\multicolumn{2}{c|}{\textbf{MNIST}} &
\multicolumn{2}{c|}{\textbf{EMNIST}} \\

\cline{2-9}

\textbf{Utilized} &
\textbf{Acc.{(\%)}} & \textbf{Loss} &
\textbf{Acc.{(\%)}} & \textbf{Loss} &
\textbf{Acc.{(\%)}} & \textbf{Loss} &
\textbf{Acc.{(\%)}} & \textbf{Loss} \\



\hline
10\% & \textcolor{blue}{\textbf{75.09}} & 0.5083 & 83.07 & 0.5838 & \textcolor{red}{\textbf{94.46}} & \textcolor{ForestGreen}{\textbf{0.2874}} & 83.66 & \textcolor{purple}{\textbf{0.6001}} \\ \hline

20\% & \textcolor{blue}{\textbf{80.54}} & 0.4943 & 83.68 & \textcolor{purple}{\textbf{0.5730}} & \textcolor{red}{\textbf{95.00}} & \textcolor{ForestGreen}{\textbf{0.2769}} & 85.12 & 0.5421 \\ \hline

30\% & \textcolor{blue}{\textbf{83.72}} & 0.3934 & 85.28 & \textcolor{purple}{\textbf{0.4551}} & \textcolor{red}{\textbf{96.10}} & \textcolor{ForestGreen}{\textbf{0.2590}} & 85.44 & 0.4420 \\ \hline

40\% & \textcolor{blue}{\textbf{85.50}} & 0.4347 & 86.28 & \textcolor{purple}{\textbf{0.4768}} & \textcolor{red}{\textbf{96.12}} & \textcolor{ForestGreen}{\textbf{0.2382}} & 86.86 & 0.4530 \\ \hline

50\% & \textcolor{blue}{\textbf{86.47}} & \textcolor{purple}{\textbf{0.4356}} & 86.75 & 0.3926 & \textcolor{red}{\textbf{96.35}} & \textcolor{ForestGreen}{\textbf{0.2347}} & 86.47 & 0.3949 \\ \hline

60\% & 87.87 & \textcolor{purple}{\textbf{0.4344}} & \textcolor{blue}{\textbf{87.08}} & 0.4261 & \textcolor{red}{\textbf{96.70}} & \textcolor{ForestGreen}{\textbf{0.2730}} & 87.25 & 0.4231 \\ \hline

70\% & \textcolor{blue}{\textbf{86.47}} & 0.3414 & 87.24 & \textcolor{purple}{\textbf{0.4127}} & \textcolor{red}{\textbf{96.88}} & \textcolor{ForestGreen}{\textbf{0.1931}} & \textcolor{blue}{\textbf{86.47}} & 0.33949 \\ \hline

80\% & \textcolor{blue}{\textbf{87.54}} & 0.3272 & \textcolor{blue}{\textbf{87.54}} & \textcolor{purple}{\textbf{0.4097}} & \textcolor{red}{\textbf{97.01}} & \textcolor{ForestGreen}{\textbf{0.2128}} & 87.55 & 0.3898 \\ \hline

90\% & \textcolor{blue}{\textbf{87.18}} & \textcolor{ForestGreen}{\textbf{0.3225}} & \textcolor{blue}{\textbf{87.18}} & \textcolor{purple}{\textbf{0.3927}} & \textcolor{red}{\textbf{97.09}} & 0.2012 & 87.31 & 0.3924 \\ \hline

\end{tabular}
\end{table}

\section{Reproducibility, Data-Experiment, Ablation Study and Representation Analysis}
\label{sec:Reproducibility_data-experiment_representation_analysis}

The dataset utilization analysis in Table ~\ref{tab:dataset_utilization} indicates that the proposed SimAM-HVQC exhibits robust learning characteristics, 
achieving reasonable performance with limited training data while benefiting from increased dataset availability. 

The ablation results of  Table ~\ref{tab:ablation} verify that the inclusion of the parameter-free SimAM attention mechanism 
contributes positively to model effectiveness, leading to improved accuracy and reduced 
classification loss across multiple benchmark datasets.

\begin{figure}[htbp]
    \centering
    \includegraphics[width=\columnwidth]{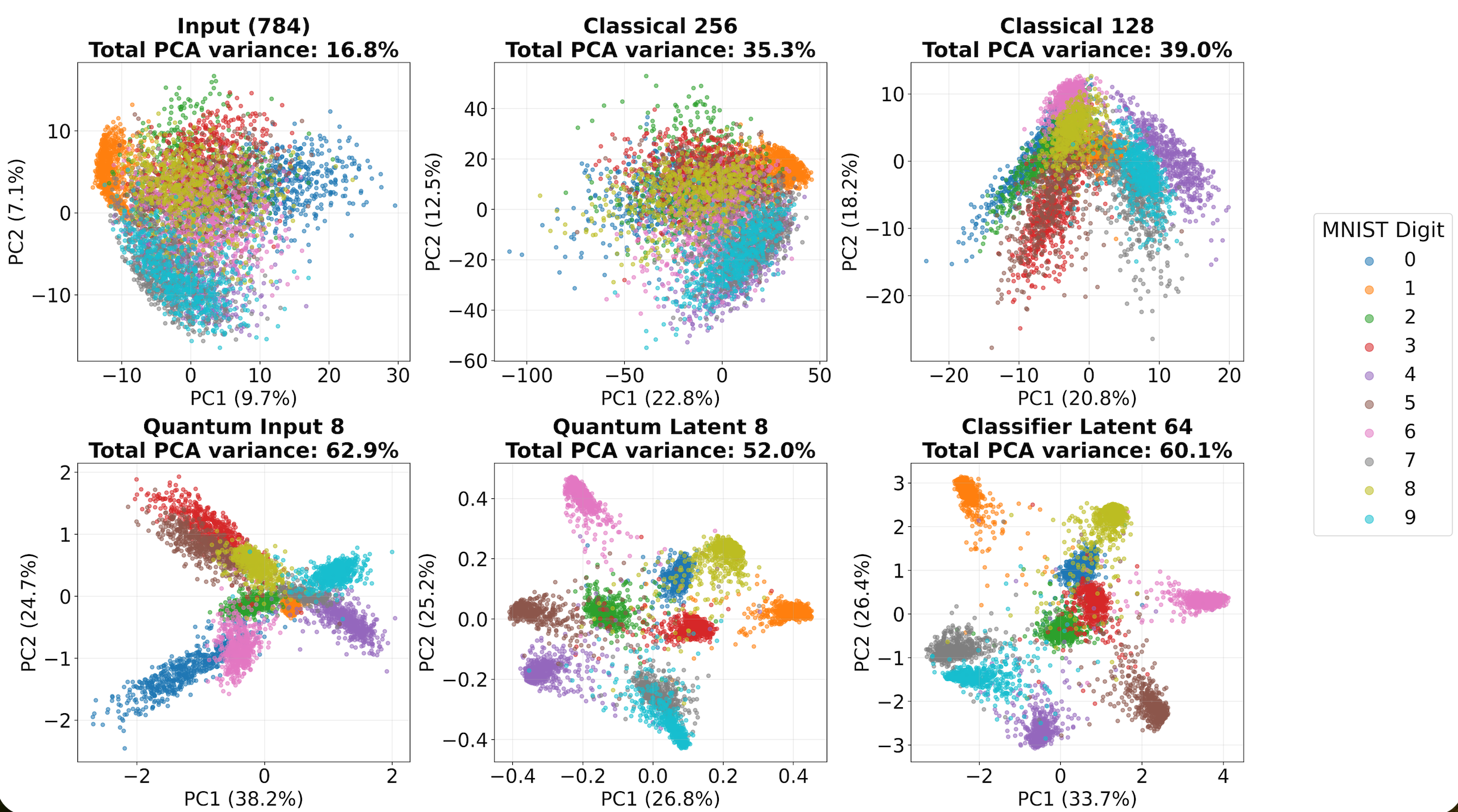}
    \caption{PCA of latent representations across the hybrid quantum-classical network. Each panel projects a layer's feature space onto its top two principal components (variance explained on axes), colored by digit label. Class separability increases with depth, with the quantum layer and final classifier layer achieving the clearest separation despite low dimensionality.}
    \label{fig:pca_architecture_flow_latent_representation}
\end{figure}

Fig.~\ref{fig:pca_architecture_flow_latent_representation} 
shows the evolution of latent representations across SimAM-HVQC. 
The input space exhibits substantial class overlap 
(16.8\% PCA variance), whereas intermediate and 
final representations become increasingly structured. 
Notably, the 8-dimensional quantum representations retain 
high projected variance (52.0--62.9\%) 
and clear class organization despite substantial 
dimensionality reduction. 
The classifier latent space exhibits the 
most distinct class clusters.
\begin{table}[t]
\centering
\caption{Efficiency Assessment of the Proposed SIM-HVQC on CPU}
\label{tab:efficiency}
\scriptsize
\begin{tabular}{|l|r|}
\hline
\textbf{Metric} & \textbf{SIM-HVQC} \\
\hline

Parameters & 236,258 \\

Qubits & 8 \\

Quantum Layers & 6 \\

Quantum Parameters & 144 \\

Test Samples & 2,000 \\

Inference Time (s) & 1.6249 \\

Latency (ms/sample) & 0.8124 \\

Throughput (samples/s) & 1230.85 \\
\hline
\end{tabular}
\end{table}

\begin{figure}[htbp]
    \centering
    \includegraphics[width=\columnwidth]{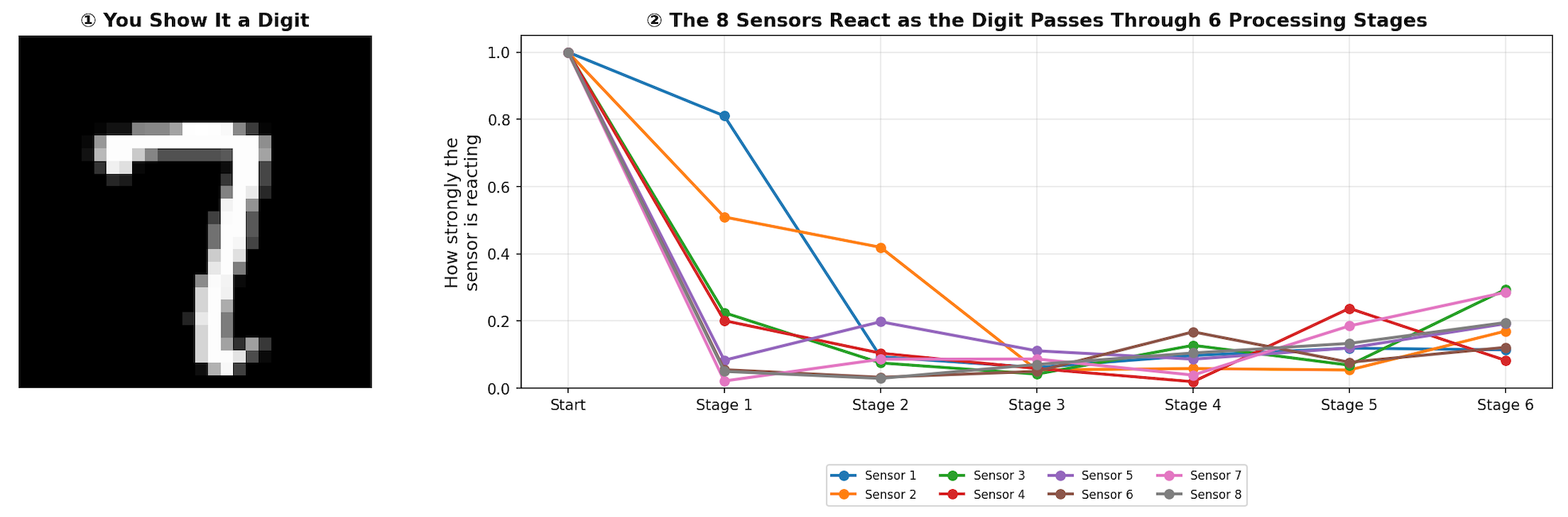}
    \caption{Shows that the eight quantum measurements progressively 
    develop distinct response patterns across the six 
    variational layers, indicating specialization of 
    quantum features and formation of a discriminative 
    quantum representation for classification.
    }
    \label{fig:qbit_measurement}
\end{figure}

\begin{table}[htbp]
\caption{Ablation Study of Sim-HVQC With and Without SimAM}
\label{tab:ablation}
\centering
\scriptsize
\begin{tabular}{|l|c|c|c|c|}
\hline
Dataset & \multicolumn{2}{c|}{SIM-HVQC} & \multicolumn{2}{c|}{No SIM-HVQC} \\
\cline{2-5}
& Accuracy (\%) & Loss & Accuracy (\%) & Loss \\
\hline
MNIST
& \textbf{97.57 $\pm$ 0.05}
& \textbf{0.1437 $\pm$ 0.0139}
& 97.48 $\pm$ 0.18
& 0.1638 $\pm$ 0.0054 \\
\hline
Fashion-MNIST
& \textbf{87.98 $\pm$ 0.08}
& \textbf{0.3345 $\pm$ 0.0132}
& 87.51 $\pm$ 0.11
& 0.4424 $\pm$ 0.0218 \\
\hline
EMNIST
& \textbf{80.16 $\pm$ 0.17}
& \textbf{0.7212 $\pm$ 0.0236}
& 78.99 $\pm$ 0.06
& 0.8921 $\pm$ 0.0302 \\
\hline
KMNIST
& 88.20 $\pm$ 0.63
& \textbf{0.3544 $\pm$ 0.0283}
& \textbf{88.23 $\pm$ 0.57}
& 0.9284 $\pm$ 0.0593 \\
\hline
\end{tabular}
\end{table}

\begin{figure}[htbp]
    \centering
    \includegraphics[width=\columnwidth]{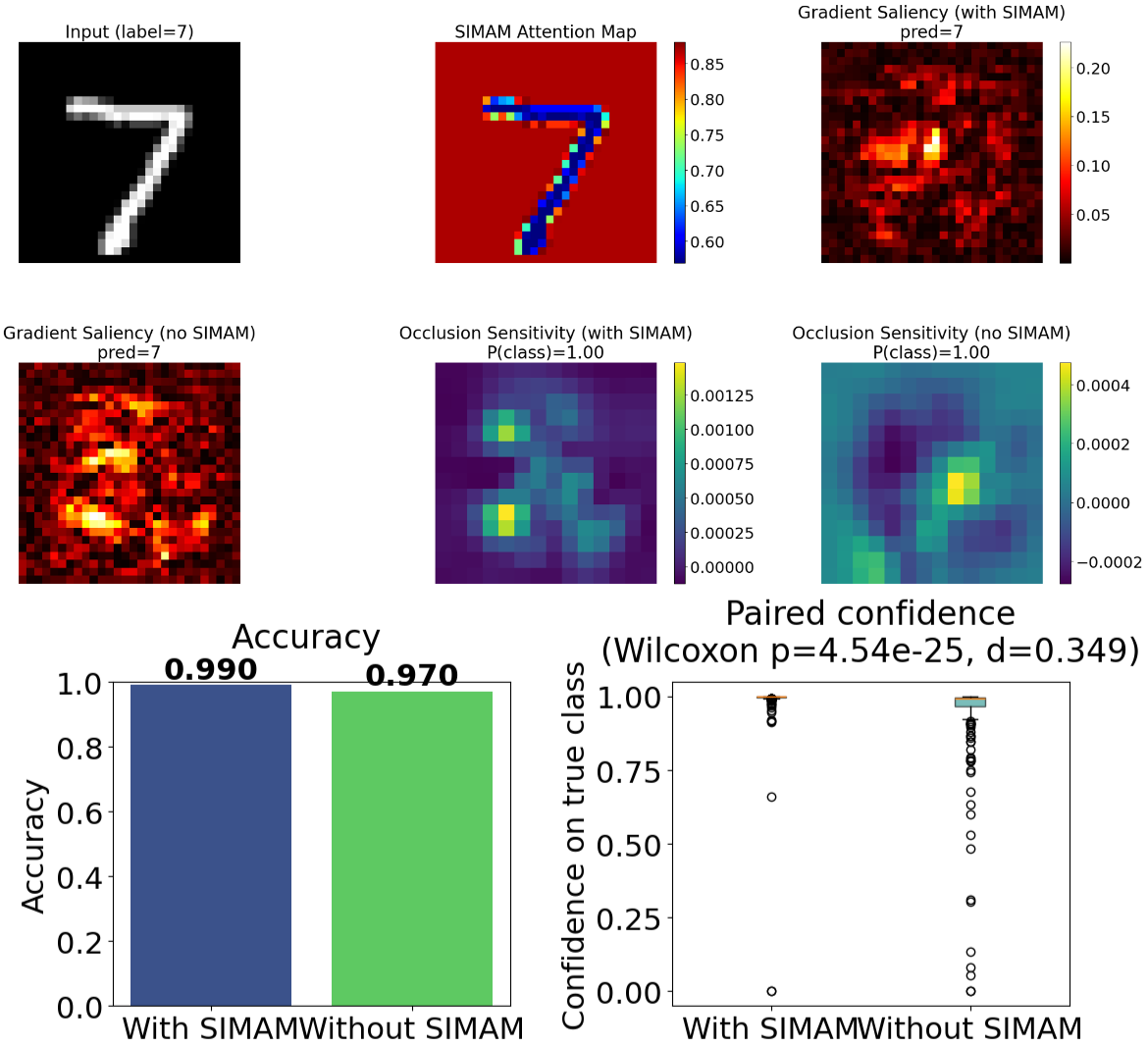}
    \caption{
        Top Figure: Attention, saliency, and occlusion maps illustrate feature importance for a sample input.
        Bottom Figure: The dataset-level comparison for 'n=200' sample test-dataset shows improved accuracy and significantly higher prediction confidence when SimAM is enabled (Wilcoxon $p=4.54\times10^{-25}$, Cohen's $d=0.349$).
    }
    \label{fig:sim_ablation_withoutablation_accuracy_stats_significant}
\end{figure}

Occlusion-based qubit sensitivity analysis illustrates the image regions (Fig.~\ref{fig:occulsion_6}) that most strongly influence individual quantum measurements. Distinct qubits exhibit sensitivity to different portions of the input pattern, suggesting that the variational quantum circuit captures complementary feature representations from the compressed latent embedding.
The test result shows of the bottom panel Figure 
~\ref{fig:sim_ablation_withoutablation_accuracy_stats_significant} 
that enabling SimAM produces a statistically significant 
improvement in prediction confidence (and related performance) 
compared to the model without SimAM.
SimAM-HVQC maintains computational efficiency (Table ~\ref{tab:efficiency}) despite incorporating quantum processing. 
The low quantum parameter count together with sub-millisecond inference latency suggests that the quantum component introduces limited computational overhead.

Evolution of quantum feature representations across the variational circuit. 
Fig.~\ref{fig:qbit_measurement} shows the layer-wise response magnitude of the eight qubits, revealing distinct feature-specialization patterns. 
The lower-left panel of Fig.~\ref{fig:qbit_measurement} illustrates the growth of inter-qubit correlations throughout the circuit depth, while the lower-right panel presents the final Pauli-Z measurement outputs that form the quantum feature vector supplied to the classifier.

\section{Conclusion}
\label{sec:conclusion}
In conclusion, the proposed SimAM-HVQC demonstrated competitive and 
promising performance across multiple standard benchmark datasets 
while maintaining parameter efficiency and supporting reproducibility. Efficiency
and interpretability analyses highlights its potential as a 
practical hybrid quantum-classical framework for representation learning and visual recognition tasks.

\bibliographystyle{IEEEtran}
\bibliography{references}

\end{document}